\documentclass{article}
\usepackage{latexsym}
\usepackage{color,cite,graphicx}
\usepackage{latexsym,amssymb,epsf} 

\newcommand{\bfzero}{{\bf 0}}
\newcommand{\bfA}{{\bf A}}
\newcommand{\bfc}{{\bf c}}
\newcommand{\bfE}{{\bf E}}
\newcommand{\bfJ}{{\bf J}}
\newcommand{\bfn}{{\bf n}}
\newcommand{\bfp}{{\bf p}}
\newcommand{\bfP}{{\bf P}}
\newcommand{\bfx}{{\bf x}}
\newcommand{\bfrho}{{\mbox{\boldmath $\rho$}}}
\newcommand{\bfsigma}{{\mbox{\boldmath $\sigma$}}}
\newcommand{\bfSigma}{{\mbox{\boldmath $\Sigma$}}}
\newcommand{\calE}{{\cal E}}
\newcommand{\calJ}{{\cal J}}
\newcommand{\calL}{{\cal L}}
\newcommand{\calN}{{\cal N}}
\newcommand{\calP}{{\cal P}}
\newcommand{\calZ}{{\cal Z}}
\newcommand{\bfxt}{\left(\bfx,t\right)}
\newcommand{\bfxtp}{\left(\bfx+\bfc_i,t\right)}

\newcommand{\bfk}{{\bf k}}

\def\bfsigma{\mbox{\boldmath $\sigma$}}

\newcommand{\bge}{\begin{equation}}
\newcommand{\ee}{\end{equation}}
\newcommand{\bgc}{\begin{center}}
\newcommand{\ec}{\end{center}}
\newcommand{\bgea}{\begin{eqnarray}}
\newcommand{\eea}{\end{eqnarray}}
\newcommand{\bgeas}{\begin{eqnarray*}}
\newcommand{\eeas}{\end{eqnarray*}}

\newcommand{\hcc}{H_{\mbox{\scriptsize cc}}}

\begin{document}

\title{
  \bf A three-dimensional lattice gas model for amphiphilic fluid dynamics
}

\author{
  Bruce M. Boghosian,\\
  {\footnotesize Center for Computational Science,
    Boston University,}\\
  {\footnotesize 3 Cummington Street, Boston,Massachusetts
    02215, U.S.A.}\\
  {\footnotesize{\tt bruceb@bu.edu}}\\[0.3cm]
  Peter V. Coveney,\\
  {\footnotesize Centre for Computational Science,
    Queen Mary and Westfield College,}\\
  {\footnotesize University of London, Mile End Road,}\\
  {\footnotesize London E1 4NS, U.K.}\\
  {\footnotesize{\tt p.v.coveney@qmw.ac.uk}}\\[0.3cm]
  and\\[0.3cm]
  Peter J. Love\\
  {\footnotesize Theoretical Physics, Department of Physics,
    University of Oxford,}\\
  {\footnotesize 1 Keble Road, Oxford, OX1 3NP, UK}\\
  {\footnotesize{\tt p.love1@physics.oxford.ac.uk}}
}

\maketitle

\newpage

\begin{abstract}
We describe a three-dimensional hydrodynamic lattice-gas model of
amphiphilic fluids. This model of the non-equilibrium properties of
oil-water-surfactant systems, which is a non-trivial extension of an
earlier two-dimensional realisation due to Boghosian, Coveney and Emerton~\cite{bib:bce}, can be studied
effectively only when it is implemented using high-performance computing
and visualisation techniques. We describe essential aspects of the
model's theoretical basis and computer implementation, and report on the phenomenological
properties of the model which confirm that it correctly captures binary
oil-water and surfactant-water behaviour, as well as the complex phase
behaviour of ternary amphiphilic fluids.
\end{abstract}

\newpage

\section{Introduction}
\label{sec:intro}

Oil and water are immiscible fluids under normal conditions of
temperature and pressure. Their phase separation behaviour in binary
mixtures has been extensively studied both experimentally and
theoretically, and has become a testbed for many fluid simulation
methods~\cite{bib:bray,bib:jury}. However, the addition of amphiphile (or
surfactant) to these systems produces much more complex behaviour. A
general review of the equilibrium phase behaviour of these fluids and
theoretical methods for studying them was provided by Gompper and
Schick~\cite{bib:gs2}. 
The complex phase behaviour of binary and ternary amphiphilic fluids arises
as a result of the physical and chemical properties of amphiphilic
molecules. Amphiphile molecules possess a hydrophylic
head and a hydrophobic tail. As a result it is usually
energetically favourable for the amphiphile molecules to be adsorbed at
oil-water interfaces, effectively reducing the interfacial tension.
Complex morphologies, termed mesophases, occur in both binary
(surfactant and water or oil) and ternary amphiphilic systems. In
general these structures depend on the concentration and chemical nature
of the surfactant molecules (length of hydrocarbon tail, size of head
group, and so on) as well as on the temperature.  The equilibrium phase
diagrams for these systems have been obtained from experimental
investigation~\cite{bib:kshks}. Of considerable interest are the {\it
microemulsion} phases; an example is the very low surface tension which
exist between the ``middle'' microemulsion phase and the bulk oil and water
phases~\cite{bib:gs2}, but there are many other fascinating phenomena
including the viscoelastic properties of wormlike micelles and the
formation of vesicles~\cite{bib:gs2,bib:cw}. Their intrinsic complexity
and wide application make these systems appropriate for
detailed scientific inquiry.  Although the equilibrium phase behaviour
of these systems is well understood, relatively little work has been
done on their non-equilibrium properties. Much of the work which has
addressed these dynamic properties has been based on molecular dynamics
methods~\cite{bib:mdpl,bib:mdshe,bib:mdmlc}.  However, such atomistic approaches are too
computationally demanding to allow investigation of the important
large-time dynamics and extended spatial structures of these
systems. Indeed, a large part of the fascination of amphiphilic fluids is
related to the dependence of their macroscopic properties on the underlying molecular
and mesoscopic dynamics, which calls for the development of techniques
which can efficiently bridge the length and time scale gaps from
micro to macro.

In the present paper we report on the formulation and implementation of
a three-dimensional hydrodynamic lattice gas model, which has been extensively studied
in two dimensions~\cite{bib:bce,bib:em1,bib:em2,bib:em3,bib:em4}. Compared with
molecular dynamics the computational simplicity of lattice gases makes
them an ideal method for modelling complex fluid behaviour from the
mesoscale upwards.  They have been used extensively for modelling
hydrodynamics since Frisch, Hasslacher, and Pomeau~\cite{bib:fchc}, and
Wolfram~\cite{bib:w4} showed that it is possible to simulate the
incompressible Navier-Stokes equations using discrete boolean elements
on a lattice. Rothman and Keller subsequently generalised
the basic lattice gas method to allow simulation of immiscible fluids~\cite{bib:rk},
and we have used their model as the starting point for our own work.

Notwithstanding the simplifications engendered by invoking the
lattice-gas paradigm, 
the simulation of the non-equilibrium behavior of ternary
amphiphilic fluids in three dimensions is a highly demanding area of
computational science; indeed, the results presented in this paper have
been made possible only by the recent availability of sufficiently
powerful parallel computing architectures, as well as sophisticated
visualisation methods.

The purpose of the present paper is to describe the implementation of
our three-dimensional model, and to establish its validity. In
particular, we show that our model
can reproduce the well known features of amphiphilic fluid
phenomenology.  In Sections~\ref{sec:model} and \ref{sec:amph_hamil} we
describe our model, emphasising in particular the differences between
the 2D and 3D lattice-gas realisations, and briefly describe the
computational requirements of the work.  Section~\ref{sec:algy} outlines
the basic structure of the algorithm, while Section~\ref{sec:couple}
specifies the coupling constants which are used in our simulations.
The results of the simulations are presented in
Section~\ref{sec:simul}. These simulations demonstrate the ability of
our model to represent binary immiscible behaviour, binary
water-surfactant self-assembly and ternary amphiphilic behaviour. 
Finally we close the paper with some conclusions in
Section~\ref{sec:conclusions}.

\section{Amphiphilic Lattice-Gas Dynamics}
\label{sec:model}

Our lattice-gas model of amphiphilic fluids consists of three different
species of particles moving about on a $D$-dimensional lattice $\calL$
in discrete time steps.  The three species are the two immiscible fluids
(oil and water) denoted by red ($R$) and blue ($B$) colours,
respectively, and the amphiphile $A$.  Lattice-gas particles can have
velocities $\bfc_i$, where $1\leq i\leq b$, and $b$ is the number of
velocities per site.  We shall measure discrete time in units of one
lattice update, so that a particle emerging from a collision at site
$\bfx$ and time $t$ with velocity $\bfc_i$ will stream to site
$\bfx+\bfc_i$ where it may undergo the next collision.  The $\bfc_i$
must thus be integer multiples of the lattice vectors; it is also
possible to have $\bfc_i=\bfzero$ for some $i$ to allow for ``rest
particles'' with zero speed.

We let $n^\alpha_i\bfxt\in\{0,1\}$ denote the presence ($1$) or absence
($0$) of a particle of species $\alpha\in\{R,B,A\}$ with velocity
$\bfc_i$, at lattice site $\bfx\in\calL$ and time step $t$.  The
$n^\alpha_i\bfxt$ are not all independent since an ``exclusion rule'' is
enforced whereby there can be no more than one particle of a given
velocity at a given lattice site at a given time.  The collection of all
$n^\alpha_i\bfxt$ for $1\leq i\leq b$ shall be denoted by
$\bfn^\alpha\bfxt$.  This is not to be confused with the total number of
particles of a given colour,
\begin{equation}
n^\alpha\bfxt\equiv\sum_{i=1}^b n^\alpha_i\bfxt.
\end{equation}
Likewise, we shall sometimes need the {\it colour charge} associated
with a given site,
\begin{equation}
q_i\bfxt\equiv n^R_i\bfxt-n^B_i\bfxt,
\end{equation}
as well as the total colour charge at a site,
\begin{equation}
q\bfxt\equiv\sum_{i=1}^b q_i\bfxt.
\end{equation}
Finally, the collection of all $\bfn^\alpha\bfxt$ for
$\alpha\in\{R,B,A\}$ will be called the {\it population state} of the
site; it is denoted by
\begin{equation}
\bfn\bfxt\in\calN
\end{equation}
where we have introduced the notation $\calN$ for the (finite) set of
all distinct population states.

In addition to the specification of the particle populations at a site,
the amphiphile particles have an orientation, denoted by
$\bfsigma_i\bfxt$.  This orientation vector, which has fixed magnitude
$\sigma$, specifies the orientation of the director of the amphiphile
particle at site $\bfx$ and time step $t$ with velocity $\bfc_i$.  (Of
course, if there is no amphiphile particle with that site, time step and
velocity, then the value of $\bfsigma_i\bfxt$ there is not defined.)  In
our work, the values of the $\bfsigma_i\bfxt$ vectors may vary
continuously~\footnote{It is also possible to construct models with
discrete set of values for the $\bfsigma_i$, but we do not consider that
possibility in this paper.} on a $(D-1)$-sphere, denoted by $S^{D-1}$;
thus, they take their values on a circle for $D=2$ and a sphere for
$D=3$.  The collection of the $b$ vectors $\bfsigma_i\bfxt$ at a given
site $\bfx$ and time step $t$ is called the {\it orientation state},
\begin{equation}
\bfSigma\bfxt
\equiv
\left(
\bfsigma_1\bfxt,
\bfsigma_2\bfxt,
\ldots
\bfsigma_b\bfxt
\right)
\in
\bigotimes^b
S^{D-1},
\end{equation}
where $\otimes^b$ denotes the $b$-fold Cartesian product.  This is not
to be confused with the {\it total director} of a site
\begin{equation}
\bfsigma\bfxt\equiv\sum_{i=1}^b\bfsigma_i\bfxt.
\end{equation}
We shall also find it useful to define the following scalar director
field
\begin{equation}
S\bfxt\equiv\sum_{i=1}^b\bfc_i\cdot\bfsigma_i\bfxt.
\label{eq:sdef}
\end{equation}
Together, the population state and the orientation state completely
specify (in fact, as noted above, they overspecify) the {\it total
state} of the site,
\begin{equation}
s\equiv(\bfn,\bfSigma),
\end{equation}
where we have omitted the site and time-step specification for brevity.

The time evolution of the system is an alternation between a streaming
or {\it propagation} step and a {\it collision step}.  In the first of
these, the particles move in the direction of their velocity vectors to
new lattice sites.  This is described mathematically by the replacements
\begin{equation}
n^\alpha_i\left(\bfx+\bfc_i,t+1\right)
\leftarrow
n^\alpha_i\bfxt
\label{eq:propp}
\end{equation}
\begin{equation}
\bfsigma_i\left(\bfx+\bfc_i,t+1\right)
\leftarrow
\bfsigma_i\bfxt,
\label{eq:propo}
\end{equation}
for all $\bfx\in\calL$, $1\leq i\leq b$ and $\alpha\in\{ R,B,A\}$.  That
is, particles with velocity $\bfc_i$ simply move from point $\bfx$ to
point $\bfx+\bfc_i$ in one time step.

In the collision step, the newly arrived particles interact, resulting
in new momenta and directors.  The collisional change in the state of a
lattice site $\bfx$ is required to conserve the mass of each species
present
\begin{equation}
\rho^\alpha\bfxt\equiv\sum_i^b n^\alpha_i\bfxt,
\end{equation}
as well as the $D$-dimensional momentum vector
\begin{equation}
\bfp\bfxt\equiv\sum_\alpha\sum_i^b\bfc_i n^\alpha_i\bfxt
\end{equation}
(where we have assumed for simplicity that the particles all carry unit
mass).  Thus, the set $\calN$ of population states at each site is
partitioned into {\it equivalence classes} of population states having
the same values of these conserved quantities.  For a site with given
masses $\bfrho\equiv\left(\rho^R,\rho^B,\rho^A\right)$ and momentum
$\bfp$, we denote the set of allowed population states by
$E(\bfrho,\bfp)\subset\calN$.  Since the conservation laws do not
restrict the orientations of the directors, the set of allowed total
states is then
\begin{equation}
\calE(\bfrho,\bfp)=E(\bfrho,\bfp)\bigotimes^b S^{D-1}.
\end{equation}

Given a precollision total state $s\in E(\bfrho,\bfp)$ with masses
$\bfrho$ and momentum $\bfp$, the postcollision total state $s'$ must
belong to the set
\begin{equation}
s'=\left(\bfn',\bfSigma'\right)
\in \calE(\bfrho(s),\bfp(s)).
\end{equation}
Henceforth, primed quantities are understood always to refer to the
postcollision state.  In our model, $s'$ is sampled from a probability
density $\calP(s')$, sometimes equivalently denoted
$\calP(\bfn',\bfSigma')$, imposed upon this set.  We assume that the
characteristic time for collisional and orientational relaxation is
sufficiently fast in comparison to that of the propagation that we can
model this probability density as the Gibbsian equilibrium corresponding
to a Hamiltonian function; that is
\begin{equation}
\calP(s')
=
\frac{1}{\calZ}\exp\left[-\beta H(s')\right], \label{eq:beta_defn}
\end{equation}
where $\beta$ is an inverse temperature, $H(s')$ is the energy
associated with collision outcome $s'$, and $\calZ$ is the
equivalence-class partition function,
\begin{equation}
\calZ (\bfrho,\bfp,\beta)
\equiv
\sum_{\bfn'\in E(\bfrho,\bfp)}
\int d\bfSigma'\;\exp\left[-\beta H(s')\right],
\end{equation}
and we have defined the measure on the set of orientational states
\begin{equation}
\int d\bfSigma\equiv\bigotimes_{i=1}^b\int d\bfsigma_i.
\end{equation}
In practice, we sample $s'=\left(\bfn',\bfSigma'\right)$ by first
sampling the postcollision population state $\bfn'$ from the reduced
probability density
\begin{equation}
P\left(\bfn'\right)
=
\int d\bfSigma'\;\calP(s').
\label{eq:redp}
\end{equation}
We then sample the postcollision orientation state by sampling the $b$
orientations $\bfsigma_i' $ from each of
\begin{equation}
\pi_i\left(\bfsigma'_i\right) = \prod_{j\neq i}^b\int
d\bfsigma'_j\;\calP\left(\bfn',\bfSigma'\right).
\label{eq:redo}
\end{equation}
for $1\leq i\leq b$; these are, as we shall see, independent
distributions, so the $b$ samples may each be taken without regard for
the other outcomes.  This completes the collision process.

\section{Local Amphiphilic Lattice-Gas Hamiltonian}
\label{sec:amph_hamil}

It remains to specify the local Hamiltonian used in the collision
outcome selection process.  Such a Hamiltonian has been derived and
described in detail for the two-dimensional version of the
model~\cite{bib:bce}, and we use the same one here~\footnote{Actually,
the forms given for the various fields in reference~\cite{bib:bce} are
somewhat more general than those given here.  In this presentation we
restrict ourselves to nearest neighbour interactions for simplicity.}.
It is
\begin{equation}
H(s') =
\bfJ\cdot (\alpha\bfE+\mu\bfP) +
\bfsigma'\cdot (\epsilon\bfE+\zeta\bfP) +
\calJ : (\epsilon\calE+\zeta\calP)+{\delta \over 2}
{{{\bf v}({\bf x},t)}^{2}},
\label{eq:hamil}
\end{equation}
where we have introduced the {\it colour flux} vector of an outgoing
state
\begin{equation}
\bfJ\bfxt\equiv\sum_{i=1}^b\bfc_i q'_i\bfxt,
\label{eq:cflux}
\end{equation}
the {\it dipolar flux} tensor of an outgoing state
\begin{equation}
\calJ\bfxt\equiv\sum_{i=1}^b\bfc_i\bfsigma'_i\bfxt,
\end{equation}
the {\it colour field} vector
\begin{equation}
\bfE\bfxt\equiv\sum_{i=1}^b\bfc_i q\bfxtp,
\label{eq:bfEdef}
\end{equation}
the {\it dipolar field} vector
\begin{equation}
\bfP\bfxt\equiv-\sum_{i=1}^b\bfc_i S\bfxtp,
\end{equation}
the {\it colour field gradient} tensor
\begin{equation}
\calE\bfxt\equiv\sum_{i=1}^b\bfc_i\bfE\bfxtp,
\end{equation}
the {\it dipolar field gradient} tensor~\footnote{Note that this
definition differs from that used in the reference~\cite{bib:bce}.  The
two definitions can be reconciled by a redefinition of the coupling
constants.}
\begin{equation}
\calP\bfxt\equiv-\sum_{i=1}^b\bfc_i\bfc_i S\bfxtp,
\label{eq:calPdef}
\end{equation}
and the kinetic energy of the particles at a site.
\begin{equation}
{\delta \over 2}\left|{\bf v}({\bf x},t)\right|^2,
\end{equation}
where ${\bf v}$ is the average velocity of all particles at a site, the
mass of the particles is taken as unity, and $\alpha$, $\mu$,
$\epsilon$, $\zeta$ and $\delta$ are coupling constants.

We note that the change in kinetic energy was not included by Rothman
and Keller in their immiscible lattice gas model. We include it here for
completeness, and set $\delta =1.0$; although it will not affect the
equilibrium properties of the model, it may impact the non-equilibrium
properties. It should be further noted that the inverse temperature-like
parameter $\beta$ is not related in the conventional way to the kinetic
energy. For a discussion of the intoduction of this parameter into
lattice gases to reproduce critical behaviour we refer the reader to the
original work by Chan and Liang~\cite{bib:chli}.

Eqs.~(\ref{eq:hamil} - \ref{eq:calPdef}) were derived by assuming that
there is an interaction potential between colour charges, and that the
directors are like ``colour dipoles'' in this context~\cite{bib:bce}.
The terms containing $\alpha$ models the interaction of colour charges
with surrounding colour charges as in the original Rothman-Keller
model~\cite{bib:rk}; those containing $\mu$ model the interaction of
colour charges with surrounding colour dipoles; those containing
$\epsilon$ model the interaction of colour dipoles with surrounding
colour charges (alignment of surfactant molecules across oil-water
interfaces); those containing $\zeta$ model the interaction of colour
dipoles with surrounding colour dipoles (interfacial bending energy or
``stiffness'').

Note that the field quantities depend only on the precollision state,
whereas the flux quantities depend on the postcollision state.  Thus,
the fields may be computed once at every site, just after the
propagation step.  The fluxes, on the other hand, must be computed for
each possible outgoing state.  It follows that the Hamiltonian may be
written in the form
\begin{equation}
H(s')=H_0(\bfn')+\sum_{i=1}^b n^{A\prime}_i\bfA_i\cdot\bfsigma'_i,
\end{equation}
where we have defined
\begin{equation}
H_0(\bfn')\equiv
\bfJ\cdot (\alpha\bfE+\mu\bfP)
\end{equation}
and
\begin{equation}
\bfA_i\equiv\sigma
\left[
\epsilon\bfE+\zeta\bfP+(\epsilon\calE+\zeta\calP)\cdot\bfc_i
\right].
\label{eq:auxvec}
\end{equation}

The reduced probability density for $\bfn'$ is then
\begin{equation}
P(\bfn')
=
\int d\bfSigma'\calP(s')
=
\frac{\exp\left[-\beta H_0(\bfn')\right]}{\calZ}
\prod_{i=1}^b\int d\bfsigma'_i\;
\exp\left(-\beta n^{A\prime}_i\bfA_i\cdot\bfsigma_i\right),
\end{equation}
and this demands evaluation of  integrals of the form
\begin{equation}
\int d\bfsigma\;\exp\left(-n\beta\bfA\cdot\bfsigma\right).
\end{equation}

In two dimensions ($D=2$), we can adopt a polar coordinate $\theta$,
chosen so that $\theta=0$ is the direction of $\bfA$, so that this
integral becomes
\begin{equation}
\int_0^{2\pi} d\theta\;\exp\left(-n\beta\sigma |\bfA|\cos\theta\right)
=
2\pi I_0\left(n\beta\sigma |\bfA|\right),
\end{equation}
where $I_0$ is the modified Bessel function of the first
kind~\cite{bib:bce}.  So, from from Eq.~(\ref{eq:redp}), we get the
reduced probability density
\begin{equation}
P(\bfn')
=
\frac{(2\pi)^b\exp\left[-\beta H_0(\bfn')\right]}{\calZ}
\prod_{i=1}^b I_0\left(n^{A\prime}_i\beta\sigma |\bfA_i|\right)
\label{eq:rpdfa}
\end{equation}
which must be evaluated numerically for every possible outgoing state.

In three dimensions ($D=3$), we adopt spherical coordinates with polar
axis in the direction of $\bfA$, so that the integral over the
orientational degrees of freedom is of the form
\begin{equation}
\int_0^{2\pi}d\phi\int_0^\pi d\theta\;\sin\theta
\exp\left(-n\beta\sigma |\bfA|\cos\theta\right)
=4\pi
\left\{
\begin{array}{ll}
1 &
\mbox{if $n=0$}\\
\frac{\sinh\left(\beta\sigma |\bfA|\right)}{\beta\sigma |\bfA|} &
\mbox{if $n=1$,}
\end{array}
\right.
\end{equation}
where $\theta$ is the colatitude and $\phi$ is the azimuthal angle.  The
reduced probability density of Eq.~(\ref{eq:redp}) is then
\begin{equation}
P(\bfn')
=
\frac{(4\pi)^b\exp\left[-\beta H_0(\bfn')\right]}{\calZ}
\prod_{i=1}^b
\left\{
1+n^{A\prime}_i
\left[
\frac{\sinh\left(\beta\sigma |\bfA_i|\right)}{\beta\sigma |\bfA_i|}-1
\right]
\right\}.
\label{eq:rpdfb}
\end{equation}

Once the reduced probability is sampled from $P(\bfn')$, we turn our
attention to the determination of the new orientation state.  For $D=2$,
the reduced probability density for the angle $\theta'_i$ is given by
\begin{equation}
\pi_i(\theta'_i) =
\frac{\exp\left(-\beta\sigma\left|\bfA_i\right|\cos\theta'_i\right)}
{2\pi I_0\left(\beta\sigma\left|\bfA_i\right|\right)}.
\label{eq:opdfa}
\end{equation}
For $D=3$, the probability density for the colatitude $\theta'_i$ and
azimuthal angle $\phi'_i$ is then
\begin{equation}
\pi_i(\theta'_i,\phi'_i)
=
\frac{
\exp\left(-\beta\sigma\left|\bfA_i\right|\cos\theta'_i\right)
\sin\theta'_i}
{4\pi\left[\frac{\sinh\left(\beta\sigma\left|\bfA_i\right|\right)}
{\beta\sigma\left|\bfA_i\right|}\right]}.
\label{eq:opdfb}
\end{equation}

\section{Algorithmic Considerations}
\label{sec:algy}

A computer implementation of our hydrodynamic lattice-gas model requires
a choice of data representation for the population state $\bfn$ and the
orientation state $\bfSigma$.  We consider these in turn.  As noted
above, the variables $n^\alpha_i\bfxt$ are not all independent because
of the ``exclusion rule'' that only one particle of any type may have a
given velocity $i$ at a given site $\bfx$ and time step $t$.  Thus, it
is inefficient to store these variables directly.  Rather, we note that
for a given $i$, $\bfx$ and $t$ there are precisely four possibilities:
there can be a particle of type $R$, type $B$, type $A$, or nothing at
all.  These four possibilities can be encoded in two bits of information
as follows:
\begin{center}
\begin{tabular}{|l|l|l|}
\hline
High Bit & Low Bit & Description \\
\hline
0 & 0 & Nothing \\
0 & 1 & $R$ particle \\
1 & 0 & $B$ particle \\
1 & 1 & $A$ particle \\
\hline
\end{tabular}
\end{center}
Thus, the population state of a given site can be represented by $2b$
bits of information.  For $D=2$, our current implementation uses a
triangular lattice (coordination number 6) and one rest particle, so
$b=7$ and 14 bits are needed to store the population
state~\cite{bib:bce}. For $D=3$, our current implementation uses a
projected face-centered hypercubic (PFCHC) lattice (coordination number
24) and two rest particles, so $b=26$ and 52 bits are needed to store
the population state~\cite{bib:fchc,bib:cam}.  The PFCHC lattice will be
described in some detail later in this section.  For now we note that in
either case, the population state easily fits in a single integer
variable; more precisely, for $D=2$ it fits in a ``short'' integer of 16
bits, and for $D=3$ it fits in a ``double-precision'' integer of 64
bits.

The orientation state requires much more storage because we have chosen
to allow the orientation angles to be continuous~\footnote{As noted in
an earlier footnote, this choice is not thought to be essential, and it
is possible that much storage space may be saved by requiring the
orientation angles to be discrete.}.  For $D=2$ we must store the $b$
polar angles $\theta_i$, each as an IEEE-format, single precision,
floating-point variable of 32 bits.  For $D=3$ we must likewise store
the two spherical angles for each velocity for a total of $2b$
floating-point variables.  While it is true that these angular variables
are not defined unless the corresponding $n^{A\prime}_i=1$, our current
implementation provides for storage for all angles at all sites and
velocities, because the computational price of dynamically allocating
and deallocating variables was not thought to be worth the savings in
storage; for very low surfactant concentrations, this assumption may be
invalid, and so the existing code might be improved.

In a language that allows for user-defined types, such as Fortran 90,
C++ or Java, it is useful to create a type for the total state of a site
that includes both the population and orientation information.  Given
this data representation, the implementation of the propagation step is
fairly obvious.  The substitutions of Eqs.~(\ref{eq:propp}) and
(\ref{eq:propo}) are made throughout the lattice.  In Fortran 90, the
CSHIFT intrinsic accomplishes periodic shifts on arrays, and it is
natural to use this to construct a subroutine that accepts an array of
type 'total state' and performs the propagation procedure on this array
as a side effect.  The above-described representation for the population
state is somewhat inconvenient in this regard, as the bit pairs
corresponding to a particular velocity must be extracted from the
integer variable before it is shifted in that direction.  When the
propagation step is completed, the new fields $S$, $\bfE$, $\bfP$,
$\calE$ and $\calP$ are computed at each site using Eqs.~(\ref{eq:sdef})
and (\ref{eq:bfEdef}) - (\ref{eq:calPdef}).

Next, the possible outcomes for the population state are enumerated
using a lookup-table procedure.  A list of all possible outgoing states
that has been sorted according to equivalence class $E(\bfrho,\bfp)$ is
precomputed and stored.  This list is of length $2^{14}=16384$ for the
$D=2$ case.  A full list for the 52-bit population state representation
in $D=3$ would have length $2^{52}$ and this is obviously much too large
to store on any existing or contemplated computer; a method of
shortening this list will be described below.  For now, assume that such
a list could be stored, and that two other lists of the same size could
be maintained that accept the current population state $\bfn$ as a key,
and return a pointer to the initial position and the number of elements
of the equivalence class $E(\bfrho,\bfp)$ in the table of sorted states.
This makes it possible to enumerate the postcollision states $\bfn'$
that respect the required conservation laws.  Note that the length of
this list may vary from site to site.

Next, each site loops over its set $E(\bfrho,\bfp)$ of allowed
postcollision states and computes the outgoing colour flux vector $\bfJ$
and the $b$ auxiliary vectors $\bfA_i$ for each one of them, using
Eqs.~(\ref{eq:cflux}) and (\ref{eq:auxvec}) respectively.  These are
then used to compute the reduced probability density $P(\bfn')$, given
by Eq.~(\ref{eq:rpdfa}) for $D=2$, or Eq.~(\ref{eq:rpdfb}) for $D=3$.
Given these probabilities, a final population state $\bfn'$ is sampled.

Once $\bfn'$ is known, the $\bfA_i$ are recalculated for that final
state, and the final orientation angles are sampled from
Eq.~(\ref{eq:opdfa}) for $D=2$, or Eq.~(\ref{eq:opdfb}) for $D=3$.  In
the latter case, we note that $\pi_i(\theta'_i,\phi'_i)$ is independent
of $\phi'_i$ so this may simply be sampled uniformly in $(0,2\pi)$.  The
colatitude $\theta'_i$ is then found by equating its cumulative
distribution function to a random number $r$ uniformly distributed
between 0 and 1, and solving for $\theta'_i$.  The result is
\begin{equation}
\theta'_i
=
\arccos
\left\{
\frac{-1}{\beta\sigma |\bfA_i|}
\ln
\left[
re^{+\beta\sigma |\bfA_i|}+(1-r)e^{-\beta\sigma |\bfA_i|}
\right]
\right\}.
\end{equation}
For $D=2$, the sampling procedure is not so simple, because of the
integral leading to the modified Bessel function.  In this case we
proceed numerically; for small values of the parameter $\beta\sigma
|\bfA_i|$, we approximate the distribution by a Gaussian centered at its
maximum, and for small values of $\beta\sigma |\bfA_i|$ we employ
rejection sampling~\cite{bib:bce}.

It remains to describe the PFCHC lattice and our method for making the
size of the lookup table more manageable.  The face-centered hypercubic
(FCHC) lattice is a regular, self-dual lattice in four dimensions with
coordination number 24.  It can be described as all integer-valued
tetrads $(i,j,k,l)$ such that $i+j+k+l$ is even.  The motivation for
using this lattice is that it is known to yield isotropic Navier-Stokes
behaviour for a single-phase fluid~\cite{bib:fchc}.

The lattice vectors are then the 24 neighbours of the site $(0,0,0,0)$,
and these can be partitioned into a subset of eight lattice vectors
called Group 1, namely $(\pm 1,0,0,\pm 1)$ and $(0,\pm 1,\pm 1,0)$, a
subset of eight lattice vectors called Group 2, namely $(0,\pm 1,0,\pm
1)$ and $(\pm 1,0,\pm 1,0)$, and a subset of eight lattice vectors
called Group 3, namely $(0,0,\pm 1,\pm 1)$ and $(\pm 1,\pm 1,0,0)$.  The
virtue of this partition is that the sixteen lattice vectors of any two
groups lie on the corners of a regular four-dimensional hypercube, and
the eight lattice vectors of the remaining group point to the face
centers of that hypercube~\cite{bib:cam}.

The projection of the FCHC lattice to the three-dimensional PFCHC
lattice can be accomplished by simply ignoring the fourth coordinate of
the 24 vectors described above.  The partition into three groups of
eight vectors is still useful to maintain, as we shall see.  One feature
of this projection is that distinct vectors of the FCHC lattice can
project to the same vector on the PFCHC lattice; for example,
$(1,0,0,1)$ and $(1,0,0,-1)$ both project to $(1,0,0)$.  We take these
24 three-vectors as the particle velocities in our $D=3$ model, and add
two rest particles to them for a total of 26 particle velocities
($b=26$).  In our computer implementation, we append the (same) two rest
particles to each of our three groups of eight lattice vectors, to
obtain three groups of ten velocities ($b=10$) each.  The idea is then
to allow collisions within each group of ten velocities separately,
updating the state of the rest particles in all three groups whenever
they change, thereby letting the rest particles provide mass and
momentum transfer between the three groups.  The three sets of
velocities are summarized in the following table:
\vspace{1cm}

\begin{center}
\begin{tabular}{|l|l|rrr|l|l|rrr|l|l|rrr|}
\hline
Group & Lattice & \multicolumn{3}{l|}{Components}  & Group & Lattice
& \multicolumn{3}{l|}{Components} & Group & Lattice &
\multicolumn{3}{l|}{Components} \\
      & Vector  & $x$, & $y$, & $z$ & & Vector  & $x$, & $y$, & $z$ &
      & Vector  & $x$, & $y$, & $z$\\
\hline
  & $\bfc_{ 1}$ &   1,& 0,& 0 &   & $\bfc_{ 1}$ &   0,& 1,& 0 &   & $\bfc_{ 1}$
&   0,& 0,& 1\\
  & $\bfc_{ 2}$ &   1,& 0,& 0 &   & $\bfc_{ 2}$ &   0,& 1,& 0 &   & $\bfc_{ 2}$
&   0,& 0,& 1\\
  & $\bfc_{ 3}$ &  -1,& 0,& 0 &   & $\bfc_{ 3}$ &   0,&-1,& 0 &   & $\bfc_{ 3}$
&   0,& 0,&-1\\
  & $\bfc_{ 4}$ &  -1,& 0,& 0 &   & $\bfc_{ 4}$ &   0,&-1,& 0 &   & $\bfc_{ 4}$
&   0,& 0,&-1\\
1 & $\bfc_{ 5}$ &   0,& 1,& 1 & 2 & $\bfc_{ 5}$ &   1,& 0,& 1 & 3 & $\bfc_{ 5}$
&   1,& 1,& 0\\  
  & $\bfc_{ 6}$ &   0,& 1,&-1 &   & $\bfc_{ 6}$ &  -1,& 0,& 1 &   & $\bfc_{ 6}$
&   1,&-1,& 0\\
  & $\bfc_{ 7}$ &   0,&-1,& 1 &   & $\bfc_{ 7}$ &   1,& 0,&-1 &   & $\bfc_{ 7}$
&  -1,& 1,& 0\\
  & $\bfc_{ 8}$ &   0,&-1,&-1 &   & $\bfc_{ 8}$ &  -1,& 0,&-1 &   & $\bfc_{ 8}$
&  -1,&-1,& 0\\
  & $\bfc_{ 9}$ &   0,& 0,& 0 &   & $\bfc_{ 9}$ &   0,& 0,& 0 &   & $\bfc_{ 9}$
&   0,& 0,& 0\\
  & $\bfc_{10}$ &   0,& 0,& 0 &   & $\bfc_{10}$ &   0,& 0,& 0 &   & $\bfc_{10}$
&   0,& 0,& 0\\
\hline
\end{tabular}
\end{center}
\vspace{1cm}

Since two bits of information are required to represent the population
state for each velocity, a total of 20 bits will specify the state
within each group.  This results in tables of length $2^{20}=1048576$.
Since the results can be stored as single-precision (32-bit, or
four-byte) integers, the tables each require a total of 4 Mbytes of
storage.  Since there are three tables, as described above, a grand
total of 12 Mbytes are devoted to table storage.  This amount of storage
is not a significant problem on modern multiprocessor supercomputers.

Once again, the use of user-defined types can simplify the
implementation of the above decomposition.  The population state can be
made a user-defined type, constructed from three integer variables.
Subroutines for propagation, table lookup, and other collision-related
operations can then be overloaded to accommodate the new type.

Despite the advantage gained by using the hydrodynamic lattice gas
method described above, the simulation of large three dimensional
amphiphilic systems remains extremely computationally demanding. The
algorithm described above has been implemented in Fortran90, and parallelised
using the MPI (Message Passing Interface) library. Simulations for the
parameter search described in Section~\ref{sec:couple} were performed on
a 512 processsor Cray T3D at the Edinburgh Parallel Computing Centre
(EPCC). The code was then ported to SGI Origin 2000 machines at
Schlumberger Cambridge Research, the National Computational Science
Alliance (NCSA) in Illinois, and at the Oxford Supercomputing
Centre. The Cray T3D and T3E systems are Massively Parallel Processor
(MPP) machines, whereas the SGI O2000 machines have cache-coherent
Non-Uniform Memory Architecture (ccNUMA).  These two parallel
architectures are very different, and the ease of moving from one to the
other illustrates the portability of modern parallelised codes. A
performance improvement of two orders of magnitude was obtained on going
from the T3D to the O2000 machines, and all the simulations described in
Section~\ref{sec:simul} were performed on the latter platforms. Baseline
performance for a $64^{3}$ system running on 8 processors is 3.3
timesteps per minute for a binary oil-water system. Performance scales
superlinearly out to 64 processors for $64^{3}$ and $128^{3}$ systems.
Computational complexity increases as $L^{3}$ as one increases the
linear system size $L$. Currently the largest attainable system size is
$128^{3}$ due to memory limitations of the O2000 machines currently
available to us. However, $256^{3}$ and $512^{3}$ systems are attainable
within the limits of the O2000 architecture.  A port to the Cray T3E
(MPP architecture) has also been performed, and a comparison on a
processor-by-processor basis gives a performance of 30-50\%\ of the
O2000, with linear as opposed to superlinear scalability. 
However, the larger number of processors available on a typical
T3E system more than compensates for this relative fall off--in one case we had
access to a 1200-node T3E for a period of one month. A full
description of the computational aspects of this model will be
provided elsewhere.

In addition to the demanding nature of the simulations themselves,
visualising the results produced can be as--and in some aspects
more--computationally demanding. 
In particular, the generation of
geometrical information required to plot 3D isosurfaces of individual
species concentration requires RAM resources of at least 1 Gbyte. Although
visualisation is today sometimes still regarded as a subsidiary activity to
numerical simulation, we have found it absolutely vital to check that
the code was working, to
distinguish different morphoplogies and to gain intuition about the very
complex dynamics of these systems.  Visualisation of the largest and
most complex systems attainable by simulation using our hydrodynamic
lattice gas demands use of the most advanced graphics engines currently
available.

\section{Definition of Coupling Constants}
\label{sec:couple}

The Hamiltonian used to determine the collision outcomes has been
specified in Eq.~(\ref{eq:hamil}). We now describe the choice of the
coupling comstants $\alpha$, $\mu$, $\epsilon$, $\zeta$. In addition to
the terms in Eq.~(\ref{eq:hamil}) there is an additional kinetic energy
term, so that the Hamiltonian becomes:
\begin{equation}
H(s') =
\bfJ\cdot (\alpha\bfE+\mu\bfP) +
\bfsigma'\cdot (\epsilon\bfE+\zeta\bfP) +
\calJ : (\epsilon\calE+\zeta\calP)+{\delta \over 2}
\left|{\bf v}({\bf x},t)\right|^{2},
\end{equation}
where ${\bf v}$ is the average velocity of all particles at a site, and
the mass of the particles is set to unity ($\delta = 1.0$).  Our model
reduces to the Rothman-Keller model for binary immiscible fluids in the
limit $\alpha\rightarrow\infty$.  The three remaining parameters control
the surfactant interactions.  These were chosen by performing an
extensive parameter search using binary water-surfactant systems, and
measuring the structure factor for these systems to look for signs of
self-assembly.  In particular, we sought strong structure-factor peaks
indicative of spherical or wormlike micelles of a characteristic size
were sought. These simulations were performed with a surfactant:water
ratio of 1:2, well above the critical concentration for the formation of
micelles.  The physical contributions of each term in
Eq.~(\ref{eq:hamil}), and therefore the effect of varying each parameter,
is described below:
\begin{itemize}
\item{$\epsilon$ controls the interaction of outgoing dipoles with the
    surrounding colour field. In ternary phases this term will send
    surfactant to oil-water interfaces. In binary water-surfactant
    phases this interaction will tend to favour flat interfaces
    between the phases;}
\item{$\mu$ controls the interaction of outgoing coloured particles
    with the surrounding dipolar field. This term will favour the
    bending of dipoles around a central colour charge, and will be
    important in creating micellar phases;}
\item{$\zeta$ controls the surfactant-surfactant interaction. For
    positive $\zeta$ this produces an attractive, ferromagnetic
    interaction between the dipoles. This term is of limited
    importance in the formation of self-assembled phases, and was set
    to 0.5 for the simulations described below.}
\end{itemize}
The key to locating the micellar phases is to find the correct
balance between $\epsilon$ and $\mu$. Strongly peaked structure
functions were obtained for the following values of $\epsilon$ and $\mu$:
\begin{center}
$ 0.75 < \mu < 2.0 $\\
$ 0.25 < \epsilon <2.0 $
\end{center}
In order to maximise the desired behaviour of sending surfactant to
oil-water interfaces while retaining the necessary micellar binary phases we chose
a canonical set of coupling constants which are kept constant throughout
all the following simulations, except in the short vesicle study
described in Section~\ref{sec:vesicle}. The values of these constants
are:
\begin{center}
$\alpha=1.0$,  $\epsilon=2.0$,   $\mu=0.75$,   $\zeta=0.5 $
\end{center}
The temperature-like parameter, $\beta$, was held constant at 1.0 for
all of the ensuing simulations.

\section{Simulations}
\label{sec:simul}

The equilibrium properties and phase diagrams of a wide variety of real
amphiphilic fluids are well known~\cite{bib:gs2}. These phase diagrams
have many features unique to three dimensional systems. To esablish the
general validity of our model it is essential that we can reproduce the
complex phase behaviour observed in real amphiphiles. In this section we
describe some of this phenomenology, and present results of the
simulations designed to reproduce it.

\subsection{Oil-Water System}

The spinodal decomposition of immiscible fluids has been extensively
studied in two dimensions, and rather less in
three~\cite{bib:bray,bib:jury}. After a quench into the two phase region an
initially homogeneous mixture separates into relatively pure
single-phase domains.
\begin{figure}[htp]
\centering
\includegraphics[width=0.4\textwidth]{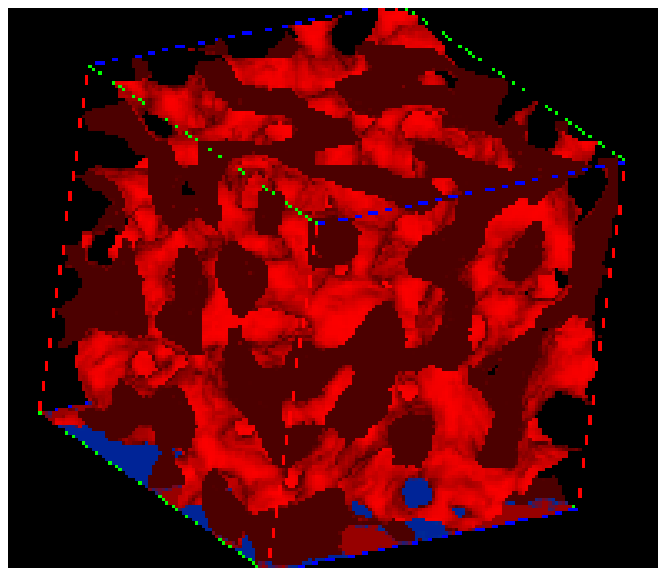}
\includegraphics[width=0.4\textwidth]{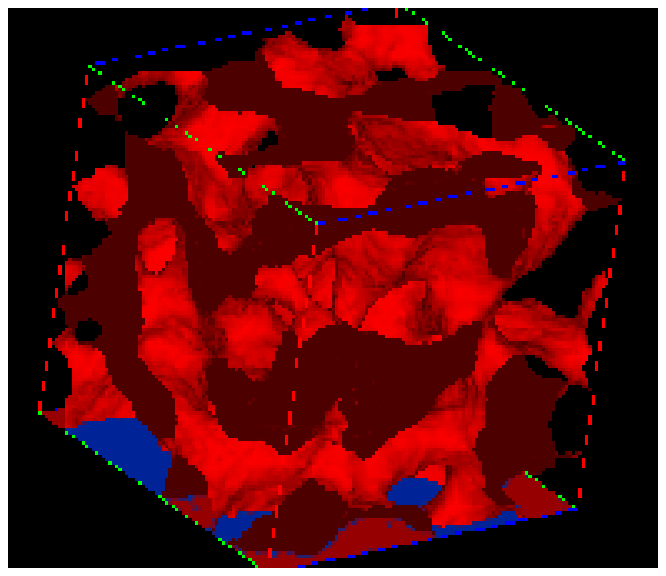}
\includegraphics[width=0.4\textwidth]{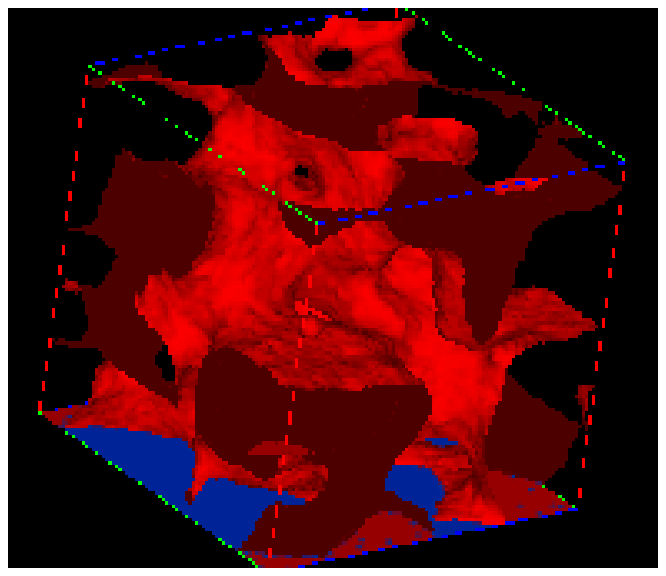}
\includegraphics[width=0.4\textwidth]{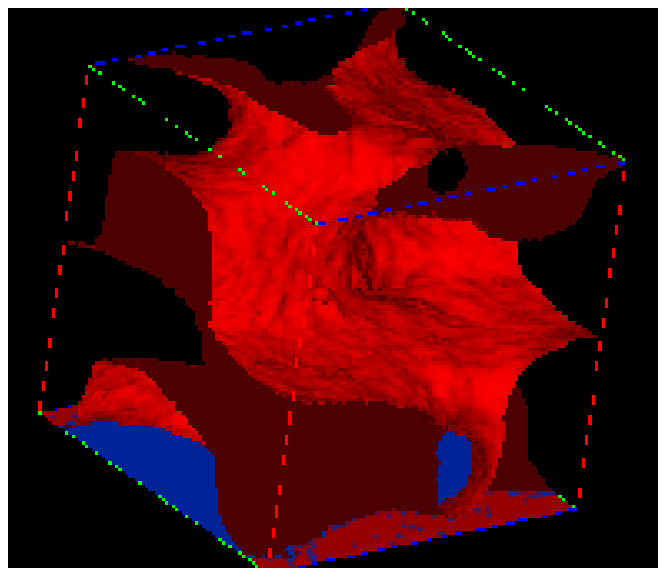}
\caption[fig:BOWS_fig1]{Binary phase separation in an oil-water system. Timesteps (clockwise
  from top left)
  200,  400, 600, 1000. Red isosurface shows interface between oil and
  water phases. System size is $64^{3}.$}\label{fig:bows_fig1}
\end{figure}

With no surfactant particles present in the system, the only term in the
local site Hamiltonian, Eq.~(\ref{eq:hamil}), that contributes
numerically to the collision process is $\alpha \Delta \hcc$.  With
parameters $\alpha=1.0$, $\beta=1.0$ we performed simulations starting
from an initial configuration in which the lattice vectors at each site
are populated randomly with oil or water particles with equal
probability (a so-called critical quench). This initial condition
corresponds to a homogenised 1:1 oil water mixture. The reduced density
(i.e. the proportion of lattice vectors at each site that contain a
particle) is $0.5$.  In this parameter regime, the model exhibits phase
separation with positive surface tension, as is evident from
Fig.~\ref{fig:bows_fig1}, which illustrates the nonequilibrium behaviour
of the immiscible lattice gas.  If left to run for a large enough time
the system would eventually reach the completely separated state of two
distinct layers of fluid.  To make a detailed comparison between the
immiscible oil-water fluid behaviour shown here and later simulations in
which we introduce surfactant to the system, we make use of {\it
spherically averaged structure functions}.  We first calculate the
three-dimensional structure factor of the colour charge, $s(\bfk, t)$,
\[
s(\bfk, t) = \left<\frac{1}{N}\left|\sum_{\bfx} (q(\bfx) - q^{av})
                   e^{i\bfk\cdot\bfx}\right|^2\right>,
\]
where $\bfk = \left(2 \pi / L \right) \left( p {\bf i} + q {\bf j} +
r{\bf k} \right)$, $p,q,r = 1,2,...,L$, q(\bfx) is the total colour
charge at a site, $q^{av}$ is the average value of the colour charge,
$L$ is the length of the system and $N = L^3$ is the number of lattice
sites in the simulation box.
\begin{figure}[htp]
\centering
\includegraphics[width=1.0\textwidth]{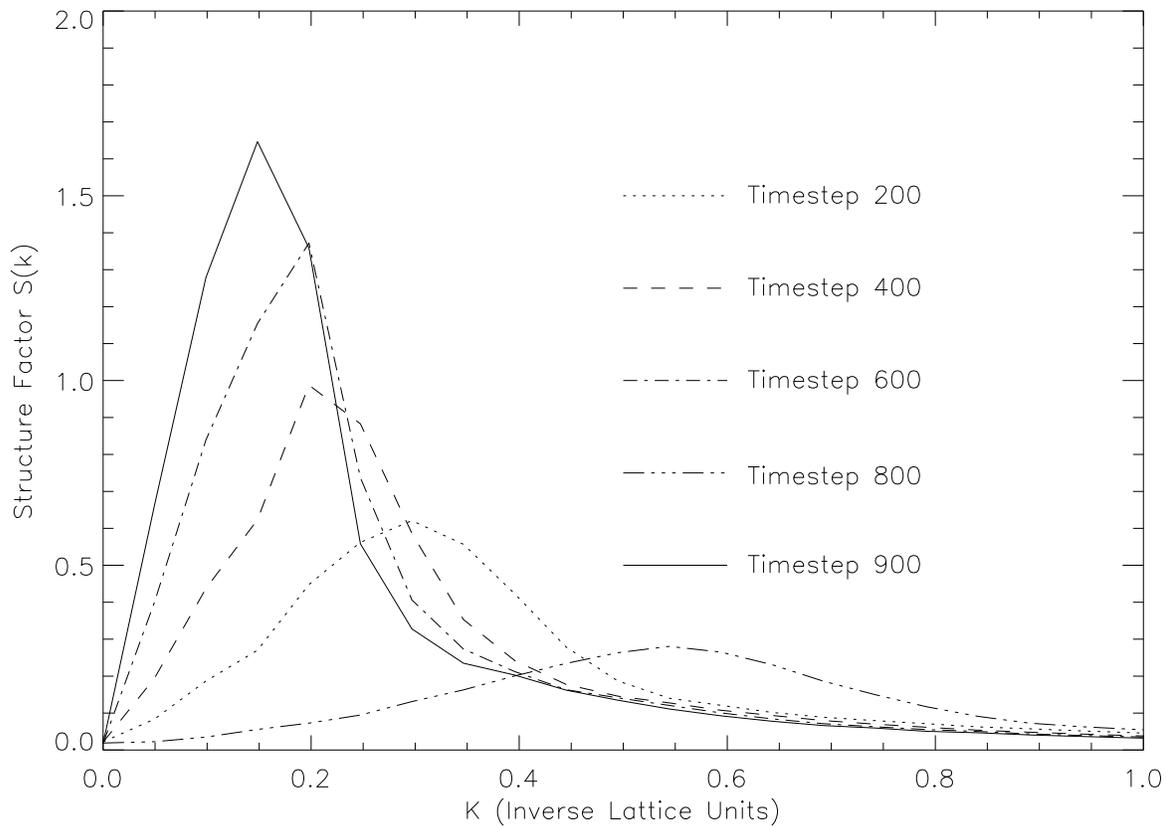}
\caption[fig:tif]{Structure factor for
timesteps 200, 400, 600, 800, 1000 in a binary phase-separating system . System size is
$128^{3}$. The structure factor measurements are averaged over ten
independent simulations.  }\label{fig:bips_sr}
\end{figure}

We actually compute the spherically averaged structure factor, given by
\bge
S(k, t) = \frac{\sum_{\hat
k}s(\bfk, t)}{\sum_{\hat k} 1},
\label{eq:castf}
\ee where $k = 2\pi n / L, n = 0,1,2,...,L$, and the sum $\sum_{\hat k}$
is over a spherical shell defined by $(n - \frac{1}{2}) \le
\frac{|\bfk|L}{2\pi} < (n + \frac{1}{2})$.  The resolution of
$S(k, t)$ depends on $k_c$, the Nyquist cutoff frequency associated with
the lattice (for a real-space sampling interval of $\Delta$ the
cut-off frequency is $1/(2\Delta)$); above this value of the frequency
there is only spurious information due to aliasing. In
our case, $k_c = (2\pi/L)n_c$, where we have chosen $n_c$ to be the
maximum possible value, which is half the lattice size.
Fig.~\ref{fig:bips_sr} shows the temporal evolution of $S(k, t)$ for the
case of two immiscible fluids.  As time increases, the peak of $S(k, t)$
shifts to progressively smaller wave numbers and its peak height increases.  This
behaviour is characteristic of domain coarsening and serves as a useful
comparison for the surfactant-containing systems described below.

\subsection{Binary amphiphilic fluid phases: from monomers to sponges.}

In two dimensions the behaviour of the binary water-amphiphile lattice
gas fluid is characterised by a {\it critical micelle concentration}
(CMC) below which the fluid is a solution of monomeric amphiphiles, and
above which the monomers form circular micelles. Further increase in
amphiphile concentration in 2D simply results in more micelles, the
micelles retaining their characteristic size.  The situation in three
dimensions is more complex. A CMC is still present for the formation of
spherical micelles, but at higher concentrations new structures appear.
As the concentration of surfactant increases the number of spherical
micelles increases, until a second critical concentration is reached,
beyond which wormlike micelles are the preferred structure.  When the
concentration of surfactant is high enough that both surfactant and
water phases percolate throughout the system a sponge phase results. We
identify this sponge phase with the $ L_3 $ phase described
by Gompper and Schick~\cite{bib:gs2}.  These concentrations have been determined in our
model for $ \beta =1 $ as:
\begin{center}
$\ 0.006\leq\rho_{spherical}  \leq 0.012 $\\
$\ 0.08 \leq \rho_{wormlike} \leq 0.25    $\\
$\ 0.25 \leq \rho_{L_{3}}$\\
\end{center}
where $\rho$ is the reduced density; that is, the fraction of lattice
sites occupied by surfactant particles.  The description of these
different regimes as distinct phases, with `critical' concentrations
separating them, may be somewhat misleading. There is a large degree of phase
coexistence. Individual monomers may join or leave a micelle in the
spherical micelle regime, and the kinetics of simple micelle formation
can be modelled theoretically on the basis of a Becker-D\"{o}ring
theory~\cite{bib:cw,bib:mdmlc}. The critical concentration for the formation of
spherical micelles is well defined in our model, with no micelles seen
below this concentration.  The formation of more complex structures,
however, appears to take place by more general (Smoluchowski-type)
agregation processes; wormlike micelles are formed from the coalesence
of spherical micelles.  Such behaviour reflects the highly dynamic
nature of our model.

\begin{figure}[htp]
\centering
\includegraphics[width=0.4\textwidth,height=0.4\textwidth]{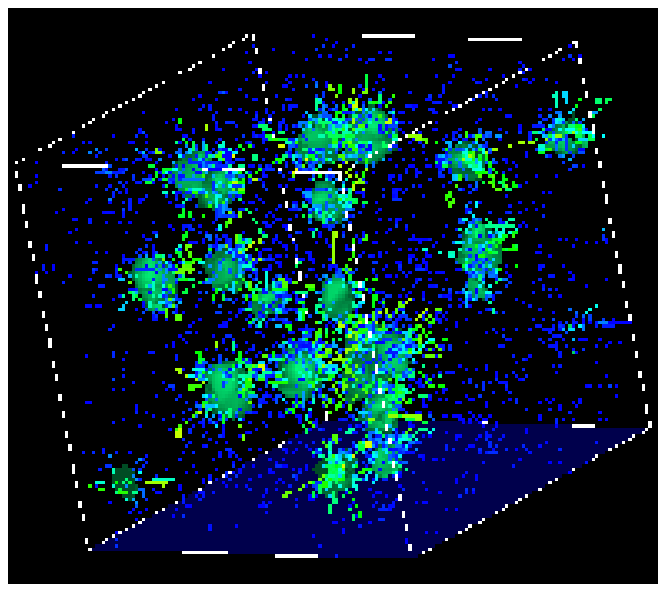}
\includegraphics[width=0.4\textwidth,height=0.4\textwidth]{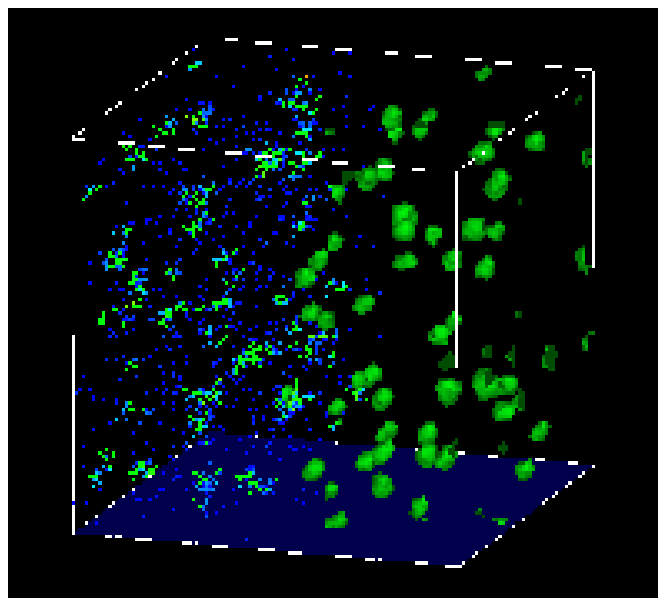}
\caption[fig:tif]{Spherical micelles in water at surfactant concentration 0.008,
timestep 1000. Left figure shows a $32^{3}$ system. Arrows represent the
average surfactant direction per site, overlaid on an isosurface of
surfactant concentration. Right figure shows a $64^{3}$ system. The left
half of this figure displays arrows of the average surfactant direction
at a site, while the right half shows an isosurface of surfactant
concentration.}\label{fig:bws_fig1}
\end{figure}

\begin{figure}[htp]
\centering
\includegraphics[width=1.0\textwidth]{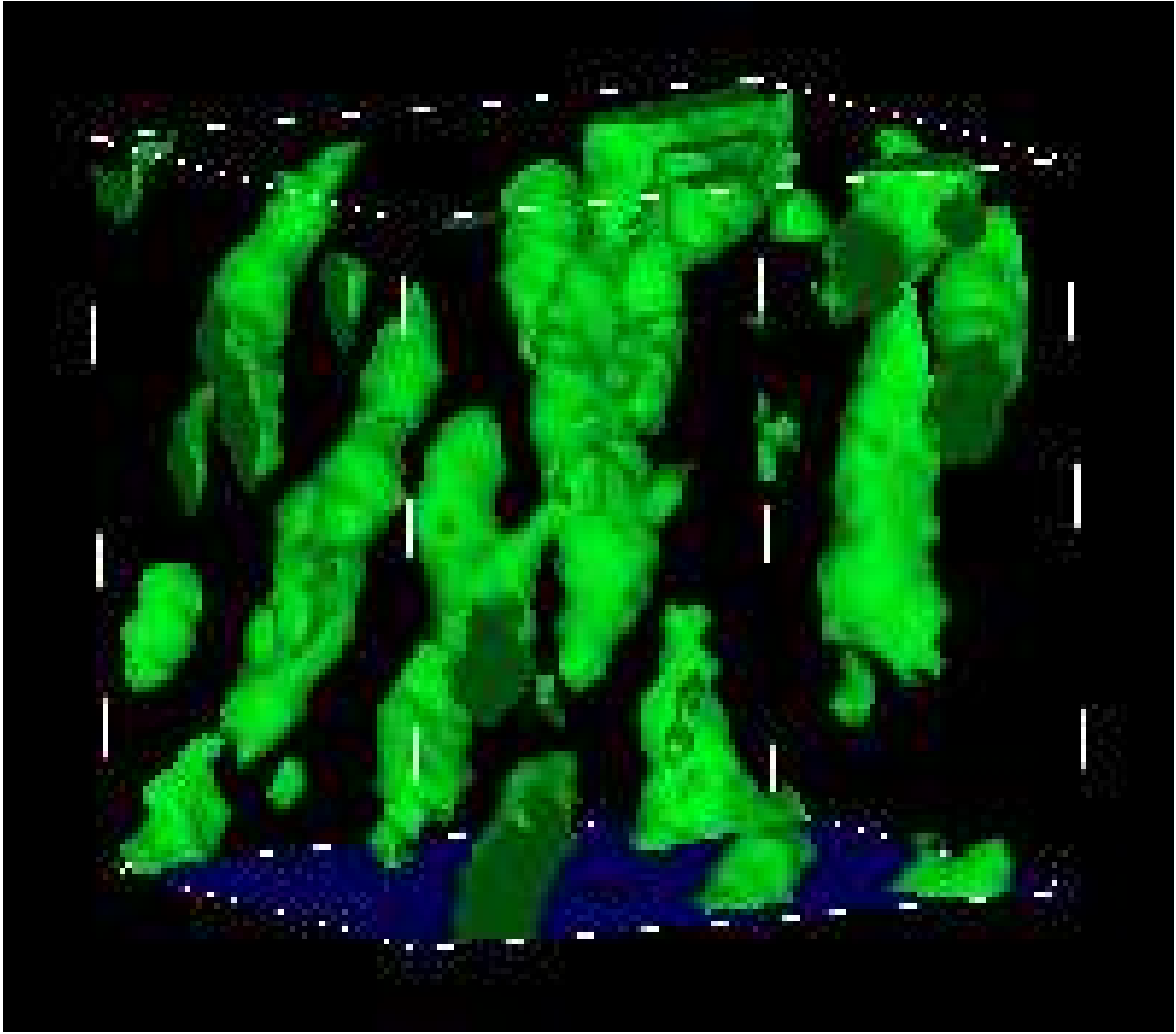}
\caption[fig:bws_fig3]{Wormlike micelles in water at a reduced surfactant
concentration of 0.1 for a $32^3$ system. The same structures are
present in larger systems, but we display this snapshot for clarity. The green
isosurface shows the surfactant concentration at a level of 5 particles
per lattice site.  }\label{fig:bws_fig3}
\end{figure}

\begin{figure}[htp]
\centering
\includegraphics[width=1.0\textwidth]{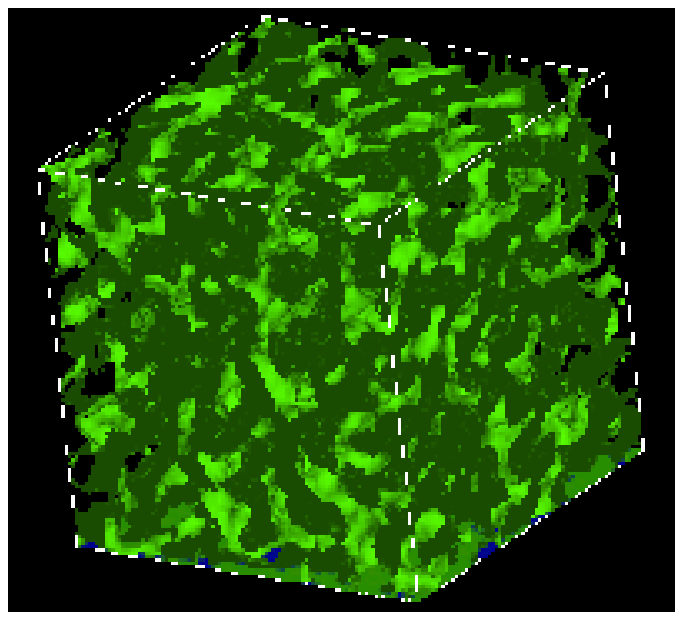}
\caption[fig:bws_fig4]{$L_{3}$ sponge phase at a reduced surfactant
concentration of 0.3. The green isosurface shows the surfactant
concentration at a level of 5 particles per site. The system size is
$64^{3}.$}\label{fig:bws_fig4}
\end{figure}

We performed simulations designed to access the monomeric, spherical
micelle, wormlike micelle and sponge phases.  All started with random
initial conditions and the coupling constants defined in
Section~\ref{sec:couple}.
\begin{figure}[htp]
\centering
\includegraphics[width=0.6\textwidth]{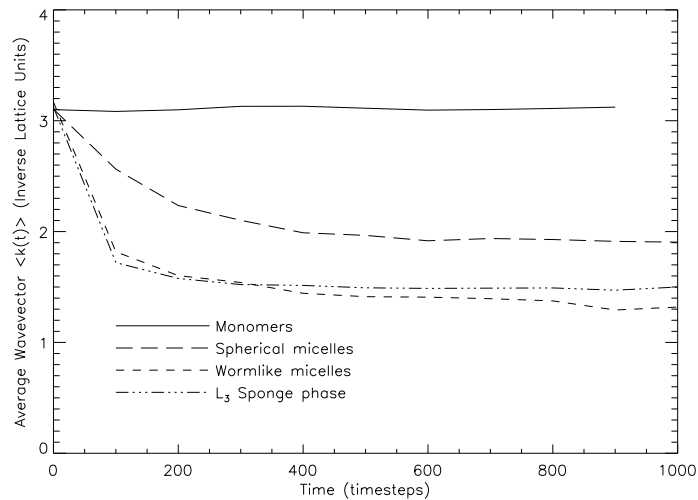}
\caption[fig:tif]{Average value of the magnitude of the 
wavevector $k$ as a function of time for varying
surfactant concentrations on a $64^{3} $ lattice. The monomeric
surfactant solution (solid line) is at reduced surfactant concentration
of 0.001; spherical micelles at reduced surfactant concentration of
0.01; wormlike micelles at reduced surfactant concentration of 0.1; and
the $L_{3}$ sponge phase at reduced surfactant concentration of
0.3}\label{fig:bws_fig2}
\end{figure}
To determine the stability of these structures, we calculated
spherically-averaged structure functions of the surfactant density.  
In Fig.~\ref{fig:bws_fig2}, we plot the temporal evolution of the
characteristic wave number,
\[
\langle k(t) \rangle =
 \frac{\sum_{k=0}^{k_c} k S(k, t)}
      {\sum_{k=0}^{k_c}   S(k, t)},
\]
the inverse of which is a measure of the average domain size.  We see
that in the low amphiphile concentration case the characteristic size
remains consistent with a random configuration of solubilised
monomers. For the cases where visualisation establishes the presence of
more complex structures there is fast initial growth of the surfactant
domains which soon levels off to some constant size; see
Figs~\ref{fig:bws_fig3} and~\ref{fig:bws_fig4}.

\subsection{Ternary phases: lamellae}

\begin{figure}[htp]
\centering
\includegraphics[width=0.4\textwidth]{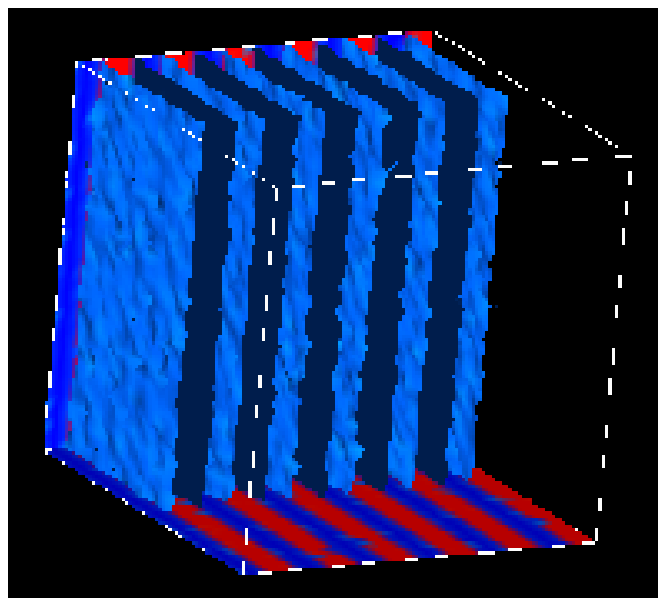}
\includegraphics[width=0.4\textwidth]{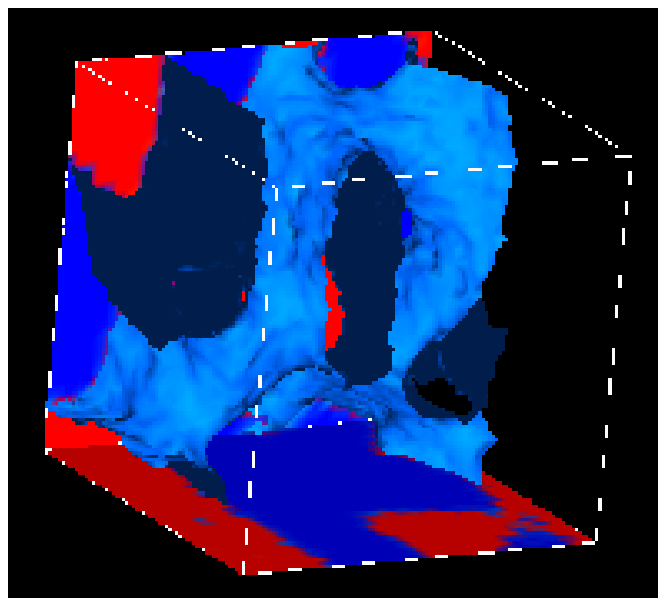}
\caption[fig:tif]{Binary oil-water lamellae at timesteps $t=0$ (left) and $t=500$ (right) for a
  $48^{3}$
  system. The red and blue colourings on the slice planes indicate oil
  and water respectively, while the blue isosurface shows the
  interface between oil and water. For clarity the isosurface is shown for
  only half of the system. The system size is $48^{3}$}\label{fig:tern_fig1}
\end{figure}

\begin{figure}[htp]
\centering
\includegraphics[width=0.4\textwidth]{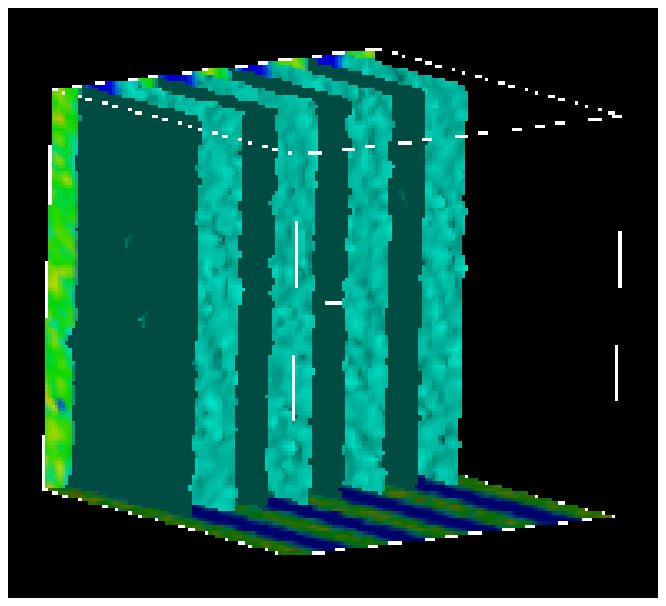}
\includegraphics[width=0.4\textwidth]{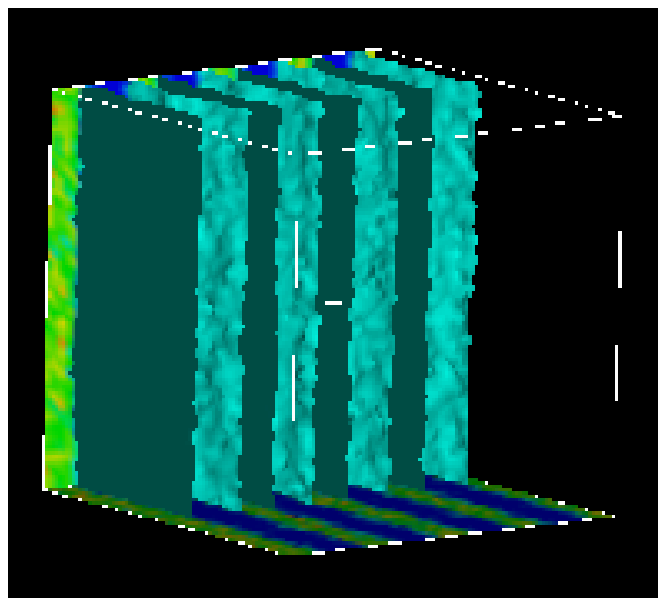}
\caption[fig:tif]{Ternary lamellae at $t=0$ (left), and $t=1000$ (right). The isosurface shows
the oil-water interface. System size is $40^{3}$.}\label{fig:tern_fig2}
\end{figure}

We first investigate the stability of a lamellar structure, which is
composed of alternating layers of oil-rich and water-rich phases
separated by a monolayer of surfactant molecules. We look at the system
with and without surfactant present in order for a comparison to be
made.  It is not clear what influence surfactant will have on such a
lamellar structure in three dimensions. We are interested in the ability
of surfactant to stabilise large areas of interface by lowering the
interfacial tension.  We set up the initial configuration of the system,
with layers of oil and water four sites wide, all sites having a reduced
density of $0.5$.  It is clear that if our model is exhibiting the
correct behaviour, then we would expect there to be a critical density
of surfactant required at the oil-water interfaces in order for the
layered structure to be stable. Consequently, we set up a simulation
with a single site wide layer of surfactant at each of the oil-water
interfaces. Snapshots from these simulations are shown in
Figs.~\ref{fig:tern_fig1} and Fig.~\ref{fig:tern_fig2}; the former is
the pure oil-water case with coupling coefficient $\alpha = 1.0$ and
inverse temperature $\beta = 1.0$, while the latter has surfactant
monolayers present.

The fluctuations present in the model enable oil and water particles from the
initially separated layers to move; they thus come under
the influence of the colour field gradients produced by other layers of
the same fluid type. For these parameter choices there is an inherent tendency
for the oil and water to act as immiscible fluids and phase separate
({\it cf.}, Fig.~\ref{fig:bows_fig1}), precisely as is shown in Fig.~\ref{fig:tern_fig1}.
Fig.~\ref{fig:tern_fig2} shows the case where surfactant monolayers coat
the oil-water interfaces. Here we see that the initial periodic
structure is stabilised despite the presence of large amounts of oil-water
interface.

\subsection{Ternary phases: oil-in-water (water-in-oil) and bicontinuous
microemulsions.}

\begin{figure}[htp]
\centering
\includegraphics[width=0.3\textwidth]{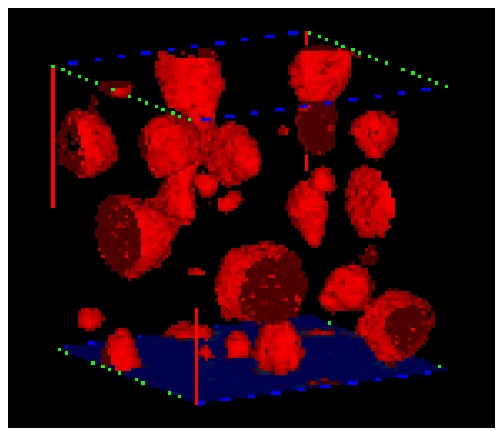}
\includegraphics[width=0.3\textwidth]{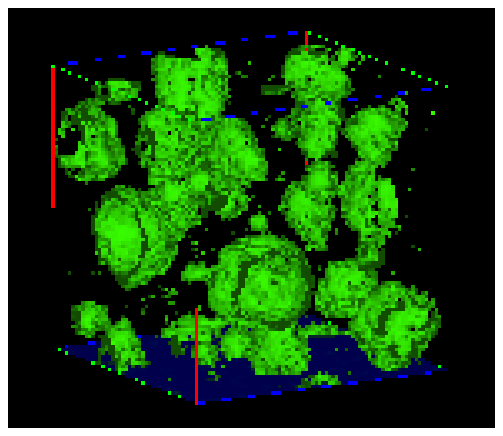}
\caption[fig:ternary3_fig1]{Oil-in-water droplet phase arising from a
   random initial condition. Oil
   (red, left) and surfactant
   (green, right) isosurfaces.} \label{fig:ternary3_fig1}
\end{figure}

The ternary behaviour of amphiphilic systems shows the same increase in
complexity between two and three dimensions that we have seen in the
binary water-surfactant systems. One simplifying feature of our model is
the symmetry between oil and water in the interactions producing the
phase behaviour.  We discuss here the two most basic of the many ternary
phases, the oil-in-water droplet and the bicontinuous {\it microemulsion}
phases.

In the oil-in-water droplet phase, oil exists as the minority phase in a
continuous background of water and surfactant. If sufficient surfactant
is present in the system the spinodal decomposition of a quenched
uniform mixture of the three components will be arrested, and the system
will reach an equilibrium state consisting of droplets of oil in water
with surfactant stabilising the oil-water interfaces.  This phase may be
regarded as a perturbation of a binary spherical micelle phase, the
micelles now being swollen with oil.  If one increases the oil
concentration, a regime is reached where both the oil and water domains
percolate throughout the system. Such a phase is the ternary extension
of the $L_{3}$ sponge phase in the binary system.

\begin{figure}[htp]
\centering
\includegraphics[width=0.7\textwidth]{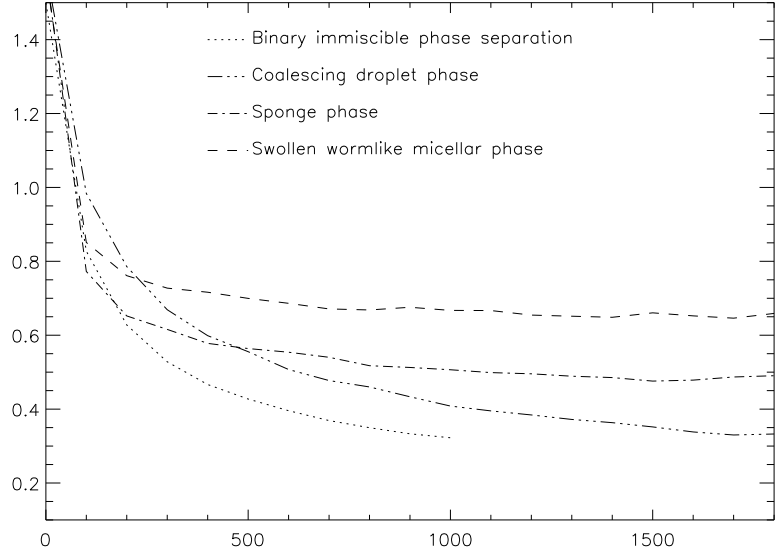}
\caption[fig:ternary3_fig2]{Structure factor mean $k$ value for binary
phase separation, ternary coalescing droplet phase, swollen wormlike
micellar phase, and bicontinuous microemulsion
phases.}\label{fig:ternary3_fig2}
\end{figure}
In both these cases, the complete separation of the oil and water
phases, which is the equilibrium state for a binary immiscible fluid, is
prevented by the presence of the surfactant.  In order to quantify this
result, and to compare with the previous binary case, we calculate the mean
$k$ value of the spherically averaged structure factor of the colour
charge, and plot the result in Fig.~\ref{fig:ternary3_fig2}.

In order to reproduce the oil-in-water droplet phase, we set up two
simulations with a random initial configuration and a reduced density of
oil of 0.05. The concentration of surfactant was varied between the two
simulations, the reduced surfactant density being 0.01 and 0.21
respectively.  The total reduced density for both simulations was 0.5;
to maintain consistency between our various simulations, we continue
to use the numerical values for the coupling constants defined in Section~\ref{sec:couple}.

\begin{figure}[htp]
\centering
\includegraphics[width=0.3\textwidth]{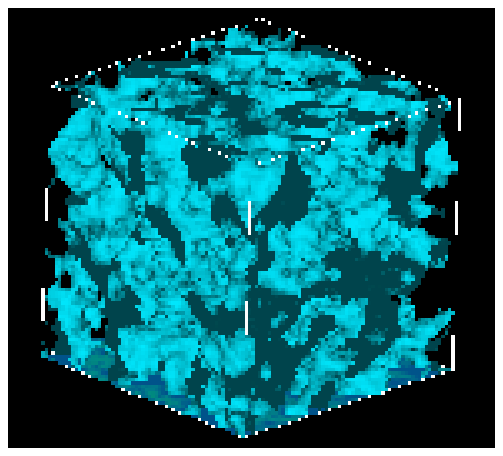}
\includegraphics[width=0.3\textwidth]{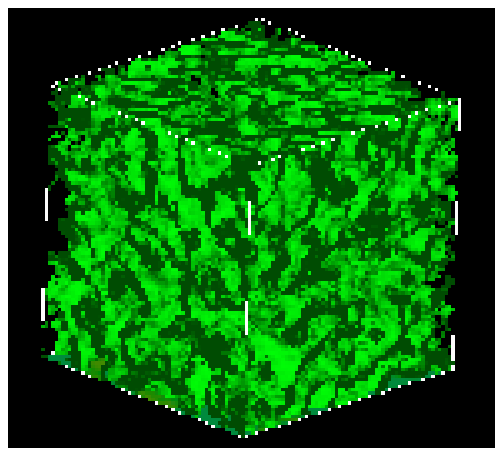}
\includegraphics[width=0.3\textwidth]{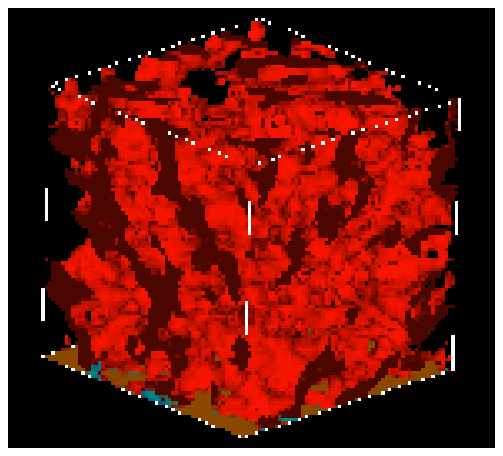}
\caption[fig:ternary3_fig3]{Bicontinuous microemulsion phase at timestep 4000.
Left: water
  isosurface; middle: surfactant isosurface; right: Oil
  isosurface. The system size is $64^{3}$. Isosurfaces display
  concentrations of water, surfactant and oil, respectively, at a
  level of five particles per site. }\label{fig:ternary3_fig3}
\end{figure}

Fig.~\ref{fig:ternary3_fig1} displays a snapshot of the simulation with
a reduced surfactant density of 0.01 at timestep 1000, displaying the
oil (red) and surfactant (green) isosurfaces. It is clear that the
surfactant is adsorbed at the interface. However,
Fig.~\ref{fig:ternary3_fig2} shows the time evolution of the average $k$
value of the spherically averaged structure factor. It is clear that
although the phase separation is slowed by the presence of surfactant,
the average domain size in the system is increasing with time.  For the
case with reduced surfactant density 0.21, one observes complete arrest
of phase separation, with wormlike oil domains remaining at a constant
size as shown by the dashed line in Fig.~\ref{fig:ternary3_fig2}. We
refer to this morphology as a swollen wormlike micellar phase. A
snapshot of this simulation taken at timestep 4000 is shown in
Fig.~\ref{fig:ternary3_fig3}.

\begin{figure}[htp]
\centering
\includegraphics[width=0.4\textwidth]{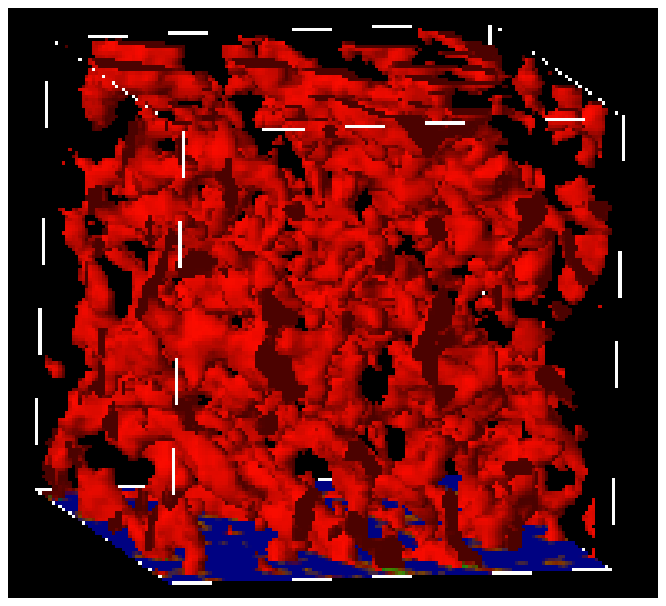}
\includegraphics[width=0.4\textwidth]{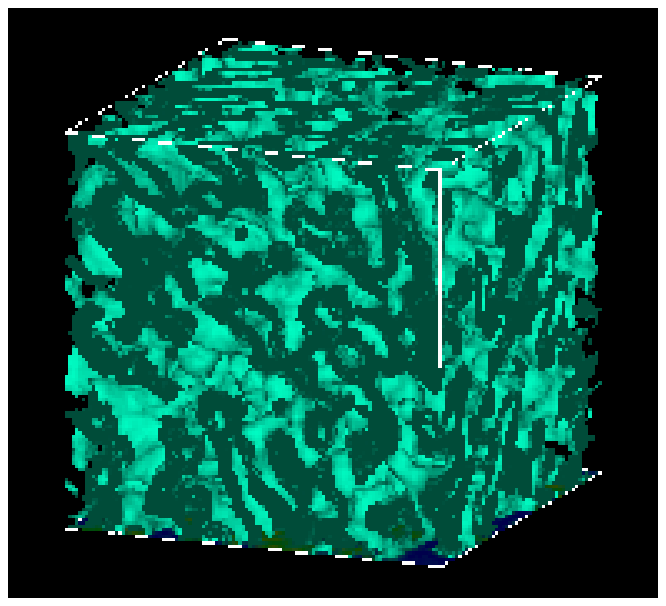}
\caption[fig:ternary3_fig4]{Stable oil-in-water droplet phase at
timestep 4000. Red and green isosurfaces show oil and surfactant
concentrations at a level of five particles per site. The system size is
$64^{3}$. }\label{fig:ternary3_fig4}
\end{figure}

To obtain the bicontinuous sponge regime within the model's phase
diagram, we simply increase the relative amount of oil present in the
system. Hence this simulation has a random initial mixture with a
reduced density $0.5$ and a $0.83:1.0:0.83$ oil-to-surfactant-to-water
ratio. Fig.~\ref{fig:ternary3_fig4} displays the results of this
simulation at timestep 4000, displaying the percolating oil (red) and
water (blue) isosurfaces, as well as the surfactant (green) isosurface.

These results show the ability of our model to correctly reproduce the
effects of surfactant on the phase separation dynamics of binary
immiscible fluids. A quantitative study of the domain growth behaviour for
both binary and ternary systems will be described in a future paper.

\subsection{Vesicles}
\label{sec:vesicle}

\begin{figure}[htp]
\centering
\includegraphics[width=0.4\textwidth]{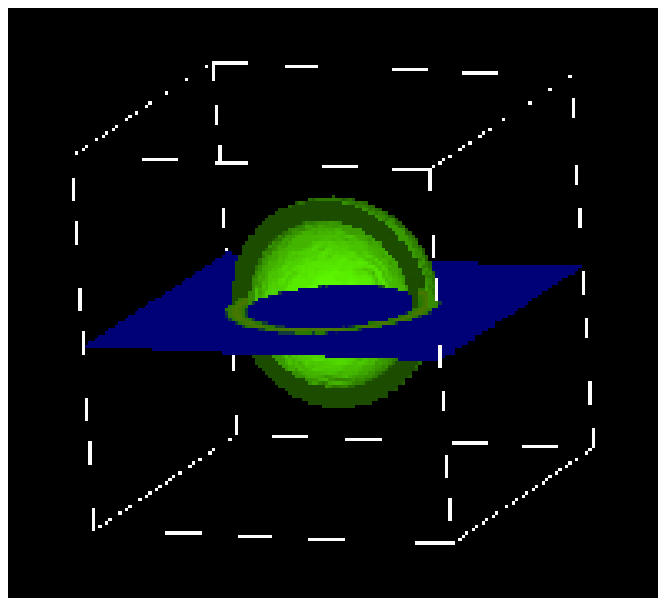}
\includegraphics[width=0.4\textwidth]{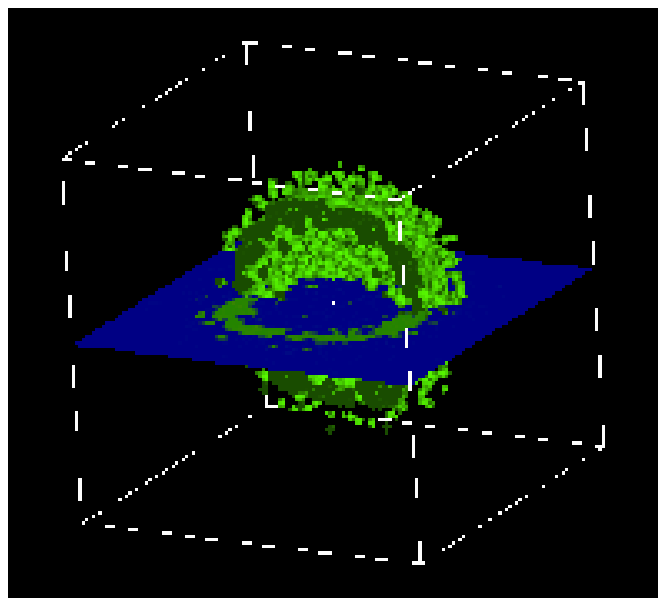}
\includegraphics[width=0.4\textwidth]{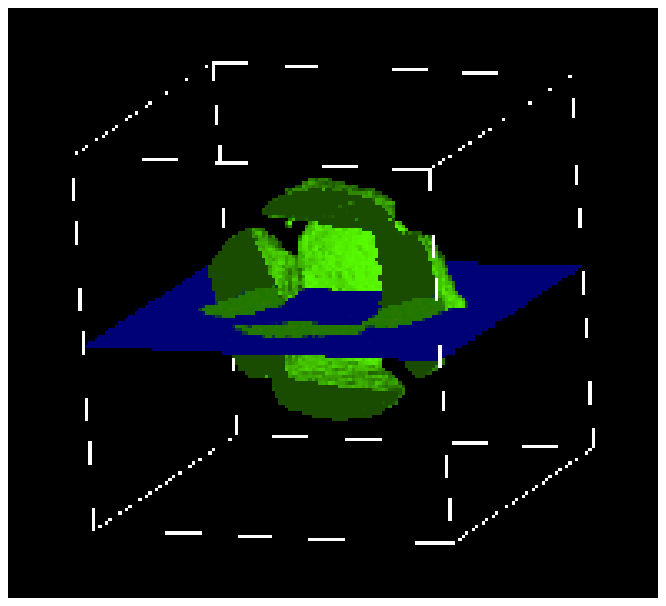}
\includegraphics[width=0.4\textwidth]{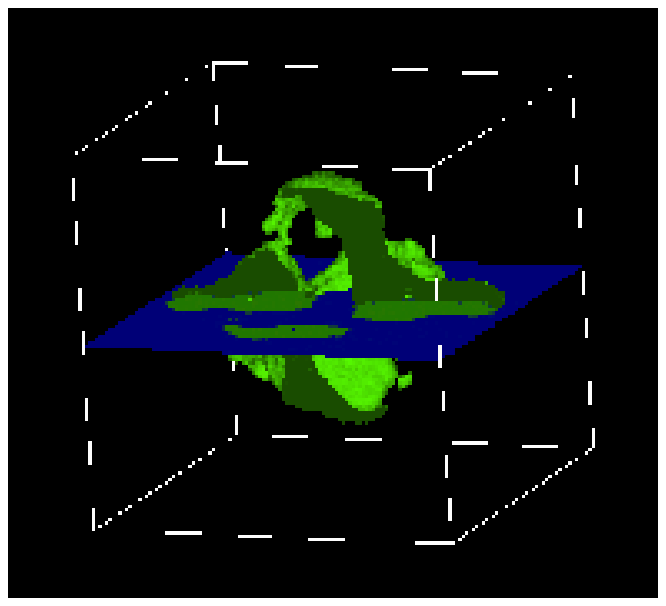}
\caption[fig:tif]{Vesicle disintegration. Clockwise from top left:
Initial condition. Simulation with $\epsilon=0.01$, $\mu=1.00$,
$\zeta=0.01 $ at timestep 200. Simulation with $\epsilon=1.00$,
$\mu=0.01$, $\zeta=0.01 $ at timestep 200. Simulation with
$\epsilon=0.01$, $\mu=0.01$, $\zeta=1.0 $ at timestep 200. All snapshots
display an isosurface showing reduced surfactant density at a level of 5
particles per site for half the vesicle. }\label{fig:vesicle_fig1}
\end{figure}

Membrane theory treatments of amphiphilic systems predict that for large
binary water surfactant systems a bilayer can reduce its energy by
closing its boundary, creating a large structure known as a
vesicle~\cite{bib:gs2}. Such structures can exist because the energy required to bend
the bilayer into a sphere is less than that required to maintain the
bilayer edge. We do not expect vesicles to self-assemble in
our current simulations, in part because such structures are usually
metastable in real life and require the input of energy, for example via shearing
or sonication. We can, however, construct such a structure as an
initial condition and study its stability, or lack thereof, using our
model.  The initial condition we have used is shown in Fig.~\ref{fig:vesicle_fig1}.
The concentration of surfactant is initially zero except within a
spherical shell, 5 sites wide and 32 sites in radius, where the reduced
density of surfactant is 1.0. All other sites contain water particles
with a reduced density of 0.5. The total system size is $128^{3}$.

For the following simulations, we have not used the canonical choice of
coupling constants described in Section~\ref{sec:couple}. Rather, in
order to study the mechanisms by which these vesicles disintegrate, we
hsve performed three separate simulations, with the following parameter choices:
\begin{center}
$\alpha=1.0$,  $\epsilon=0.01$,   $\mu=0.01$,   $\zeta=1.0 $\\
$\alpha=1.0$,  $\epsilon=0.01$,   $\mu=1.00$,   $\zeta=0.01 $\\
$\alpha=1.0$,  $\epsilon=1.00$,   $\mu=0.01$,   $\zeta=0.01 $\\
\end{center}
These three simulations should separate the effects of the three
surfactant interactions on the vesicles. We also wish to characterise
the timescale over which the vesicles are stable. In order to do this,
we performed one simulation with oil particles replacing the surfactant
particles in the initial condition, and with $\alpha=0.0$. This
essentially labels the particles starting in the vesicle region, and
allows us to track their subsequent motion.  Not suprisingly, this
simulation shows that the initial vesicle configuration is destroyed in
the absence of interactions in less than ten timesteps.  The initial
condition and state of the system at timestep 200 for the above three
simulations is shown in Fig.~\ref{fig:vesicle_fig1}.  These simulations
show that the mechanisms driving the breakup of the vesicles are phase
separation and micellisation. The vesicles are also unstable against
fluctuations in concentration which produce holes in the structure. Once
a hole has been produced it is observed to grow until the vesicle is destroyed.

\section{Conclusions}
\label{sec:conclusions}

We have extended our hydrodynamic lattice-gas model for the dynamics of
binary and ternary amphiphilic fluids from two to three dimensions. We
have shown that our model exhibits the correct 3D phenomenology using a
combination of visual and analytic techniques. Experimentally observed
self-assembling structures form in our simulations in a consistent
manner when the relative concentrations of the three components are
varied. Binary immiscible, binary amphiphilic and ternary amphiphilic
behaviour are all captured using a single set of coupling constants. We
have also shown that studies of vesicle metastability are possible
using this model with different choices of the coupling constants.
Work is currently in progress on a wide range of amphiphilic fluid
systems, including measurements of viscosity and surface tension, and the 
study of growth laws in amphiphilic self-assembly processes. Studies
of amphiphilic fluid flow in porous media, which have
previously been
performed in 2D~\cite{bib:pvcjb}, are now underway using the 3D
model~\cite{bib:pvc}.  Indeed there is a rich seam of problems related to amphiphilic
fluids which may be mined using this model. Our work confirms the
suitability of lattice gas automata for the modelling and simulation of
such complex fluid problems in both two and three dimensions.

\section*{Acknowledgements}

We are indebted to numerous people and organisations for their support
of and contributions to this work. They include Jean-Bernard Maillet and David Bailey at Schlumberger Cambridge
Research, Silicon Graphics Incorporated (particularly Bart van Bloemann
Waanders, Daron Green and Rob Jenkins), Oxford Supercomputing Centre
(particularly Jeremy Martin and Kathryn Measures), the EPSRC E7 High
Performance Computing Consortium (particularly Sujata Krishna), the
Edinburgh Parallel Computing Centre (particularly Mario Antonioletti),
and the National Computational Science Alliance in Illinois, USA. PJL would like to
thank EPSRC and Schlumberger Cambridge Research for funding his CASE
studentship award.  The collaboration between PVC and BMB was supported by NATO
grant number CRG950356.  BMB was supported in part by the United States
Air Force Office of Scientific Research under grant number
F49620-95-1-0285.

\end{document}